\documentclass[aps,prd,twocolumn,showpacs,superscriptaddress]{revtex4}
\usepackage{bm}
\newcommand{\beq}{\begin{equation}}
\newcommand{\eeq}{\end{equation}}

\begin{document}

\title{ Neutrino and antineutrino oscillations from decay of ${\rm Z}^0$ boson}
\author{I.~M.~Pavlichenkov}
\email{pavi@mbslab.kiae.ru} \affiliation{Russian Research Center
Kurchatov Institute, Moscow, 123182, Russia}

\date{\today}

\begin{abstract}
The two-particle wave function of neutrino and antineutrino after
the decay of a ${\rm Z}^0$ boson is found as a solution of an
initial value problem. The wave packet of this state involves the
coherent superposition of massive neutrinos and antineutrinos
resulting in independent oscillations of their flavor
counterparts. Possible experiments to observe oscillations of a
single lepton or both leptons from the same decay or the
independently emitted pairs $\nu_e\bar{\nu}_e$,
$\nu_\mu\bar{\nu}_\mu$ and $\nu_\tau\bar{\nu}_\tau$ are
considered.

\end{abstract}

\maketitle

The standard theories of massive neutrino oscillations consider
the time-evolution of a neutrino independently from a recoil. In
spite of the fact that almost all of these phenomenological
theories provide the canonical formula for the probability of
oscillations, some important questions arise which are discussed
in Ref.~\cite{Akh}. It has been shown in my recent work \cite{Pav}
that, in a two-particle decay, noninteracting particles, a
neutrino and a recoil, do not evolve separately due to quantum
correlation between the space-like separated objects. Because of
this the decay energy converts to the kinetic energy of the
recoil-neutrino pair. Thus, I have rigorously proved that the
neutrino mass eigenstates produced in the electron capture decay
have the same energy. The question remains: does this result still
stand for other two-body weak decays?

In this communication I apply the formalism developed in
Ref.~\cite{Pav} to the description of the decay of a ${\rm Z}^0$
boson, ${\rm Z}^0\rightarrow \nu_\alpha+\bar\nu_\alpha$, where
$\nu_\alpha$ is the neutrino with a fixed flavor
($\alpha=e,\mu,\tau$). In spite of the exotic source and the long
oscillation length compared with the Earth-Moon distance this
problem has physical interest because two entangled particles,
neutrino and antineutrino, may oscillate. This is yet another
peculiar realization of the Einstein-Podolsky-Rosen thought
experiment \cite{Ein}. The question as to whether neutrino
oscillations are observable in the invisible ${\rm Z}^0$-decay has
been considered by Bilenky and Pontecorvo \cite{Pont} and Smirnov
and Zatsepin \cite{Zat}. However, the consensus among these
authors on this subject does not exist. Let us note also that
${\rm Z}^0$-decay differs noticeably from the electron capture
decay because of the small parameter $\alpha=Q/M$ (where $Q$ is
the decay energy and $M$ is the parent particle mass) used in
Ref.~\cite{Pav} is equal to 1.

The analysis presented below is based on the solution of the
time-dependent Schr\"{o}dinger equation with initial wave function
of a decaying ${\rm Z}^0$ boson taken in the form of a finite-size
wave packet in its rest frame. The Weisskopf-Wigner theory of
spontaneous emission \cite{Wei} is used to solve the coupled
system of differential equations describing the time evolution of
the boson and the neutrino-antineutrino pair in the course of the
decay process. The wave function which describes the
spatiotemporal behavior of the pair is obtained at times
$t>1/\Gamma$, where $\Gamma$ is one half of the rate of the decay.
In the far zone approximations this function is represented as a
product of the relative and the center of mass wave functions, but
it does not factorize in the neutrino, $\mathbf{r}_{\nu}$, and
antineutrino, $\mathbf{r}_{\bar{\nu}}$, spatial coordinates. Thus
neutrino and antineutrino are entangled. It will be shown that the
coherent superposition of the neutrino and antineutrino mass
eigenstates has a fixed kinetic energy equal to the decay energy.

The decaying system is described by the Hamiltonian
\begin{equation}\label{hm}
H=\!\frac{\hat\mathbf{p}^2}{2M}+H_0+H_w,
\end{equation}
where $\hat\mathbf{p}$ is the boson momentum operator, $H_0$ is
the unperturbed Hamiltonian of the boson, neutrino and
antineutrino fields, respectively, and the weak interaction
Hamiltonian is given by
\begin{equation}\label{hmw}
H_w\!=\!\!\!\sum_{i,j,\mathbf{p},\mathbf{k}}\!\!\!\!
\frac{M_{fa}(\mathbf{k},\mathbf{p})}{\sqrt{2M}}\,U^*_{\alpha
i}U_{\alpha j} b_{\,\mathbf{p}}c^+_{i\,\mathbf{p}/2+\mathbf{k}}
d_{j\,\mathbf{p}/2-\mathbf{k}}\!+\!{\rm h.c.},
\end{equation}
where $U$ is the mixing matrix, $b$, $c$ and $d$ are the second
quantized operators of the boson, the three ($i,j=1,2,3$) massive
neutrino and antineutrino, respectively. The transition matrix
element from the initial state $a$ to the final one $f$ has the
form \vspace{-3mm}
\begin{equation}\label{mel}
M_{fa}\!=\!-\frac{i}{2}g_Z\varepsilon_\sigma \bar{u}_L(\mathbf{k}
\!+\!\mathbf{p}/2)\gamma^\sigma\scriptstyle{\frac{1}{2}}
\displaystyle(1\!-\!\gamma^5)v_R(-\mathbf{k}\!+\!\mathbf{p}/2),
\end{equation}
where $g_Z$ is the coupling constant, $u_L$ and $v_R$ are the
Dirac spinors for the left-handed neutrino and the right-handed
antineutrino, and the $Z^0$-boson transverse polarization
$\varepsilon_\sigma=-(0,1,i,0)/\sqrt{2}$ corresponds to the state
in which its spin is directed along the axis $z$ of the rest
frame.

The time-dependent perturbation theory is used to determine the
temporal evolution of a decaying state. The solution of the
Schr\"{o}dinger equation with the Hamiltonian (\ref{hm}) is sought
by using the following ansatz for the wave function
$$
\Psi(t)=\sum_\mathbf{p}A(\mathbf{p},t)b^+_\mathbf{p}|0\rangle
e^{-iE_at}
$$
\vspace{-6mm}
\begin{equation}\label{anz}
+\sum_{i,j,\mathbf{p},\mathbf{k}}\!\!\!
B_{ij}(\mathbf{p},\mathbf{k},t)c^+_{i\,\mathbf{p}/2+\mathbf{k}}
d_{j\,\mathbf{p}/2-\mathbf{k}}|0\rangle e^{-iE_{ij}t},
\end{equation}
where $|0\rangle$ is the vacuum state and the eigenvalues of the
non-perturbed Hamiltonian are
\begin{equation}\label{eigv}
E_a=M\!+\!\frac{\mathbf{p}^2}{2M},\ \
E_{ij}=\!\epsilon_i(\mathbf{k}+\mathbf{p}/2)\!+
\!\epsilon_j(\mathbf{k}-\mathbf{p}/2),
\end{equation}
where $\epsilon_i(\mathbf{k})=\sqrt{k^2+m^2_i}$ is the energy of
the massive neutrino or antineutrino with the mass $m_i$. The
coefficients $B_{ij}(\mathbf{p},\mathbf{k},t)$ can be found in
Weisskopf-Wigner approximation just as they where in
Ref.~\cite{Pav}
\begin{equation}\label{coef}
B_{ij}(\mathbf{p},\mathbf{k},t)\!=\!\frac{A_0(\mathbf{p})
M_{fa}U^*_{\alpha i}U_{\alpha j}}{\sqrt{2M}(E_{ij}-
E_a+i\Gamma)}\! \left[1\!-\!e^{i(E_{ij}-E_a)t-\Gamma t}\right],
\end{equation}
where the initial value of the coefficient $A$,
\begin{equation}\label{init}
A_0(\mathbf{p})\!=\!
(2\sqrt{\pi}d)^{3/2}\exp\left(\!-\frac{1}{2}d^2p^2\right),
\end{equation}
is the boson wave function in the momentum representation which
describes boson spatial localization before decay. The one half of
the rate of the transition ${\rm
Z}^0\!\rightarrow\!\nu_\alpha+\bar\nu_\alpha$ is given by
\begin{equation}
\Gamma=\!\frac{\pi}{2M}\!\sum_{i,j,\mathbf{k}}\!\left|U_{\alpha i}
\right|^2\left|U_{\alpha j}\right|^2
\left|M_{fa}(\mathbf{k},\mathbf{p})\right|^2\!\delta(E_{ij}-\!E_{a}),
\end{equation}
The small dependence of $\Gamma$ on the boson velocity is ignored
because $p\ll k$, and ultrarelativistic approximation reduces this
expression to a conventional value
\begin{equation}\label{gam}
\Gamma=\frac{g^2_ZM}{192\pi}=86\,{\rm MeV}.
\end{equation}

Now one can write the spatial part of the neutrino-antineutrino
wave function in the coordinate representation. With $B_{ij}$
taken from Eq.~(\ref{coef}) at times $t\!>\!1/\Gamma$, it has the
form
$$
\Psi\!=\!\frac{1}{\sqrt{2M}}\!\sum_{i,j,\mathbf{p},\mathbf{k}}
\!\frac{A_0(\mathbf{p})M_{fa}(\mathbf{k},\mathbf{p})U^*_{\alpha
i}U_{\alpha j}|\nu_i\bar{\nu}_j\rangle}
{\epsilon_i(\mathbf{k}\!+\!\mathbf{p}/2)\!+\epsilon_j(\mathbf{k}\!-
\!\mathbf{p}/2)\!-\!M\!-\!\frac{\mathbf{p}^2}{2M}\!+i\Gamma}
$$
\vspace{-3mm}
\begin{equation}\label{func}
\times\exp\!\left\{i\mathbf{k}\mathbf{r}\!+\!i\mathbf{p}\mathbf{R}-
\!i[\epsilon_i(\mathbf{k}\!+\!\mathbf{p}/2)\!+
\epsilon_j(\mathbf{k}\!-\!\mathbf{p}/2)]t\right\},
\end{equation}
where $\mathbf{r}=\mathbf{r}_\nu-\mathbf{r}_{\bar{\nu}}$ is the
coordinate of relative motion (RM) and
$\mathbf{R}=(\mathbf{r}_\nu+\mathbf{r}_{\bar{\nu}})/2$ is the
center of mass (CM) one.

This function is inconvenient to consider the entanglement and
neutrino oscillations, because it involves integration over
$\mathbf{k}$ and $\mathbf{p}$. The analytical integration over
$\mathbf{k}$ can be performed only approximately. The key
approximation is referred to as the far zone approximation, which
means that $kr\sim Mt> M/\Gamma\gg 1$. In this approximation the
main contribution to the integral over $d\Omega_{\mathbf{k}}$ is
given to those $\mathbf{k}$ close to the direction of
$\mathbf{r}$. In line with this assumption, one can put
$\mathbf{k}$ parallel $\mathbf{r}$ everywhere in the integrand of
Eq.~(\ref{func}) except in the factor
$\exp(i\mathbf{k}\mathbf{r})$. The remaining function is easily
integrated to give two terms proportional to $\exp(ikr)$ and
$\exp(-ikr)$ corresponding to outgoing and incoming spherical
waves. In the far zone, the incoming wave gives an exponentially
small contribution which can be neglected \cite{Pav1}.

For the outgoing wave, the vectors $\mathbf{k}$ and $\mathbf{r}$
are parallel and their direction is determined by the unit vector
$\mathbf{n}(\theta,\varphi)=\mathbf{r}/r$, which is the
characteristic feature of a far zone. Thus stated, the result of
the integration of Eq.~(\ref{func}) over $d\Omega_{\mathbf{k}}$ is
given by
$$
\Psi(\mathbf{R},\mathbf{r},t)\!=\!\frac{iM_{fa}(\mathbf{n})}
{(2\pi)^2\sqrt{2M}\,r} \sum_{i,j,\mathbf{p}}\!
A_0(\mathbf{p})\,e^{i\mathbf{p}\mathbf{R}}U^*_{\alpha i} U_{\alpha
j}|\nu_i\bar{\nu}_j\rangle
$$
\vspace{-4mm}
\begin{equation}\label{func1}
\times\!\!\int\limits^\infty_0
\frac{\exp\!{\left\{ikr\!-\!i\left[2\epsilon_{ij}(k)
\!+\!\mathbf{p}^2/4\epsilon_{ij}(k)\right]\!t\right\}}kdk}
{2\epsilon_{ij}(k)\!-\!M\!+\!\mathbf{p}^2/4\epsilon_{ij}(k)\!
-\!\mathbf{p}^2/2M+i\Gamma},
\end{equation}
where the small parameter $p/M$ has been used. The matrix element
(\ref{mel}) in this approximation transforms to the form
\begin{equation}\label{mel}
M_{fa}(\mathbf{n})=-\frac{g_ZM}{2\sqrt{2}}(1-\cos\theta)e^{i\varphi}.
\end{equation}
The energy $\epsilon_{ij}(k)$ depends on the mass term
$(m^2_i+m^2_j)/2$.

The remaining integral in (\ref{func1}) is calculated using the
residue method by changing the variable of integration from $k$ to
$\epsilon$ and expanding the former around the half of the decay
energy $M/2$
\begin{equation}\label{exp}
k(\epsilon)=k_{ij}+\frac{1}{v_{ij}}(\epsilon-M/2),
\end{equation}
where $k_{ij}$ and the group velocity of neutrinos,
$v_{ij}=2k_{ij}/M$, are taken at the energy $M/2$. The lower limit
of integration in (\ref{func1}) can be extended to $-\infty$
(since the contribution to the integral falls of sharply with
increasing $|\epsilon|$ owing to $M\gg\Gamma$) and the integrant
is continued analytically in complex plane $\epsilon$. In the
lowest order of the small parameter $p/M$ the pole of the
integrand is
\begin{equation}\label{pol}
\epsilon_p=(M-i\Gamma)/2,
\end{equation}
and the integral is evaluated straightforwardly
\begin{equation}\label{int}
\oint\limits_C\frac{k_{ij}(\epsilon)e^{iP(\epsilon)}d\epsilon}
{v_{ij}(\epsilon)(\epsilon -\epsilon_p)}\!=\!-\frac{2i\pi
k_{ij}}{v_{ij}}\,e^{iP_0}\Theta\!\left(2v_{ij}t\!-\!r\right).
\end{equation}
Here the contour $C$ encloses the pole (\ref{pol}). The unit step
function $\Theta$ is due to the requirement of the integral
convergence on the contour $C$. The exponent $P(\epsilon)$ is
defined as follows
\begin{equation}
P(\epsilon)=\left[k_{ij}+(\epsilon-M/2)/v_{ij}\right]r\!-\!\left(2\epsilon
+\mathbf{p}^2/4\epsilon\right)t,
\end{equation}
and its value in the pole can be transformed to the form
\begin{equation}\label{expn}
P_0=-Mt+k_{ij}r-\frac{\mathbf{p}^2t}{2M}+
\frac{i\Gamma}{2v_{ij}}\left(2v_{ij}t\!-\!r\right).
\end{equation}

Now, the wave function of relative motion can be constructed by
using Eqs.~(\ref{func1}), (\ref{int}) and (\ref{expn}). In doing
so, it is convenient to express the coupling constant $g_Z$ in
terms of $\Gamma$ (\ref{gam}). As a result the RM wave function
can be written as
\begin{equation}\label{rmf}
\psi(\mathbf{n},r,t)\!=\!F(\mathbf{n})\sum_{i,j}R_{ij}(r,t)
U^*_{\alpha i}U_{\alpha j}|\nu_i\bar{\nu}_j\rangle,
\end{equation}
where its radial part involves the functions
\begin{equation}\label{rad}
R_{ij}(r,t)\!=\!\!\frac{\sqrt{\Gamma}}{r}\exp{\left(ik_{ij}r\!-\!
\frac{2v_{ij}t\!-\!r}{D_{ij}}\right)}
\Theta\left(2v_{ij}t\!-\!r\right),
\end{equation}
depending on $r$ and $t$, and the angular one,
\begin{equation}\label{ang}
F(\mathbf{n})=\sqrt{\frac{3}{16\pi}}(1-\cos\theta)e^{i\varphi},
\end{equation}
depends on the orientation of the vector $\mathbf{r}$. The
function (\ref{rmf}) is normalized in the ultrarelativistic limit,
when $v_{ij}=1$ and $R_{ij}\!=\!R_0e^{ik_{ij}r}$, where
\begin{equation}\label{urel}
R_0(r)\!=\!\frac{\sqrt{\Gamma}}{r}\!
\exp\left[-\Gamma(t\!-\!r/2)\right]\Theta(2t\!-\!r).
\end{equation}
The function $\psi$ describes a coherent superposition of
exponential wave packets of massive neutrino-antineutrino pairs
with different sharp edges at $r=2v_{ij}t$ and widths
$D_{ij}=2v_{ij}/\Gamma\sim 10^{-13}$ cm. The later depend only on
the dynamics of the decay process and may change later due to
dispersion. To estimate such spreading effect, it is necessary to
extend the series expansion of the function $k(\epsilon)$ in
Eq.~(\ref{exp}) up to the second order in $\epsilon-M/2$. This
gives rise to an additional integral in Eq.~(\ref{func1}) as shown
in Ref.~\cite{Pav}. The calculations described in this work gives
the spreading time of the RM wave packet
\begin{equation}
t_{RM}=\frac{M^3D^2_{ij}}{8v_{ij}(m^2_i+m^2_j)}\sim 100\,{\rm s}.
\end{equation}
The time required for a neutrino or an antineutrino to propagate
the distance comparable with oscillation length, "flight time", is
10 times smaller than $t_{RM}$. This allows to use the sharp edge
wave packets for analysis of neutrino oscillations in ${\rm
Z}^0$-decay.

The integration over $d\mathbf{p}$ in Eq.~(\ref{func1}) can be
readily performed by using expressions (\ref{init}) and
(\ref{expn}). The resulting CM wave function has the standard form
for the decaying bipartite system \vspace{-4mm}
\begin{equation}
\Psi_{CM}(\mathbf{R},t)\!=\!\frac{1}{\pi^{3/4}\!
\left(d\!+\!\!\frac{it}{Md}\right)^{3/2}}
\exp{\!\!\left[\!-\frac{\mathbf{R}^2}{2d\left(d\!+\!\!\frac{it}{Md}\right)}\right]}.
\end{equation}
This is a spreading wave packet with the time-dependent width
\begin{equation}
D_R(t)=\!\left[d^2\!+\!\left(\frac{t}{Md}\right)^2\right]^{1/2}\!\!\!=\!
\left\{\begin{array}{cc} d,& t\ll t_{CM}\\
t/Md,& t\gg t_{CM} \\\end{array}\right.,
\end{equation}
where $t_{CM}=Md^2$ is its spreading time. This time is less than
the flight time if $d<10^{-3}$ cm.

The total wave function describing the evolution of the
neutrino-antineutrino pair after decay is a product of the CM and
RM parts
\begin{equation}\label{tfun}
\Psi(\mathbf{R},\mathbf{r},t)\!=\!\Psi_{CM}(\mathbf{R},t)\,
\psi(\mathbf{n},r,t)\exp(-iMt).
\end{equation}
It is seen that the energy of this state is equal to the decay
energy, $Q=M$. This is a general feature of the bipartite decay
into noninteracting fragments constrained only by momentum and
energy conservation. The wave packet
$\mid\!\psi(\mathbf{R},\mathbf{r},t)\!\mid^2$ increases with time
in a longitudinal direction (along the vector $\mathbf{r}$) with
the velocity 2 and in a transverse one with the velocity
$1/(2\pi^{3/2}Md)$. Hence, the spatiotemporal behavior of the
joint quantum state of the neutrino and the antineutrino following
$\rm{Z}^0$-decay is in agreement with the results obtained in the
theoretical studies of the electron capture decay \cite{Pav}. The
distinctive feature of our system is the coherent superposition of
the two leptons with flavor mixing, which both oscillate. To
describe this phenomenon, it is necessary to consider the
two-particle wave function (\ref{tfun}) in the observable
coordinates of the neutrino and the antineutrino. This normalized
function has the form
\begin{equation}\label{entf}
\Psi(\mathbf{r}_\nu,\mathbf{r}_{\bar{\nu}},t)\!=\!
\Psi_{CM}\!\!\left(\!\frac{\mathbf{r}_\nu\!+\!\mathbf{r}_{\bar{\nu}}}{2}
,t\!\right)\!\psi(\mathbf{n},\mathbf{r}_\nu\!-\mathbf{r}_{\bar{\nu}}
,t)\,e^{-iMt}.
\end{equation}
The function does not factorize in these variables -- a direct
indication of the spatial entanglement of leptons. Each of the
three massive neutrinos is entangled with three massive
antineutrinos.

It is instructive to consider the recoilless decay of an
infinitely heavy mother particle ($M\rightarrow\infty$). Setting
$\mathbf{p}=0$ in Eq.~(\ref{eigv}) and performing exactly the same
calculations as above with $A_0=1$, one can obtain the function of
the pair
\begin{equation}
\Psi_\infty(\mathbf{r}_\nu,\mathbf{r}_{\bar{\nu}},t)\!=\!
\delta\!\left(\frac{\mathbf{r}_\nu\!+\!\mathbf{r}_{\bar{\nu}}}{2}
\right)\!\psi(\mathbf{n},\mathbf{r}_\nu\!-\mathbf{r}_{\bar{\nu}}
,t)\,e^{-iMt}.
\end{equation}
This state is also entangled because massive neutrinos and
antineutrinos are not emitted in a momentum eigenstate due to the
fact that the interaction Hamiltonian (\ref{hmw}) commutes only
with the total momentum.

The spatial part of the neutrino-antineutrino wave packet
(\ref{entf}) for fixed $t$ is proportional to a product of the
Gaussian and exponential functions
\begin{equation}\label{pak}
\left|\Psi(\mathbf{r}_\nu,\mathbf{r}_{\bar{\nu}})\right|^2\sim
\exp{\left\{\!-\frac{(\mathbf{r}_\nu\!+\mathbf{r}_{\bar{\nu}})^2}
{4D^2_R}+2\frac{|\mathbf{r}_\nu\!-\mathbf{r}_{\bar{\nu}}|}{D}\right\}},
\end{equation}
where $D_{ij}\approx D=2/\Gamma$ in the ultrarelativistic
approximation. The probability distribution of the two-particle
coordinates depends on $r_\nu$, $r_{\bar{\nu}}$ and the angle
between these vectors. Therefore, the packet has an axially
symmetrical shape with respect to the axis passing through the CM
in the direction of the vector $\mathbf{n}$. It is easily to see
that probability reaches its peak when the vectors have equal
lengths and are located on the symmetry axis antiparallel to each
other. Hence, the daughter particles are traveling back-to-back in
the rest frame of $Z^0$ as provided by the classical picture.
However, the decay is not isotropic because of the spin
polarization of the boson. The preferred direction in which the
particles travel is determined by the factor
$|F(\mathbf{n})|^2\sim (1-\cos\theta)^2$, which is to say the
antineutrino travels mainly in the direction of the boson spin.

The structure of the neutrino-antineutrino wave packet of the
function (\ref{entf}) suggests two kinds of experiments to observe
flavor oscillations.

{\it Neutrino and antineutrino oscillations in one detector
experimen.} The first is a well-established experiment for
detecting neutrino or antineutrino oscillation. Suppose that only
a neutrino is detected regardless of the antineutrino position. In
the detection of a flavor neutrino, when the position of a
detector is scanned, such measurements allows to determine the
oscillation picture. The flavor-changing transition
$\nu_\alpha\rightarrow\nu_\beta$ is determined by the probability
density
\begin{equation}\label{no1}
\frac{dP_{\alpha\beta}}{d\mathbf{r}_\nu}=\int\!\!
d\mathbf{r}_{\bar{\nu}} \Bigr|\sum_l\!\langle\nu_l|\,U_{\beta
l}\Psi(\mathbf{r}_\nu,\mathbf{r}_{\bar{\nu}},t)\Bigl|^2.
\end{equation}
Changing the variable of integration from $\mathbf{r}_{\bar{\nu}}$
to $\mathbf{r}$, one gets the integral
\begin{equation}\label{intg}
\int\!\frac{d\mathbf{r}}
{r^2}\exp{\left[-\frac{(\mathbf{r}\!+\!2\mathbf{r}_\nu)^2}
{4D^2_R}\!+\!\frac{2r}{D}\right]}|F(\theta\varphi)|^2
W_{\alpha\beta}(r) \Theta(2t\!-\!r),
\end{equation}
where
\begin{equation}\label{prob}
W_{\alpha\beta}(r)=\delta_{\alpha\beta}-4\sum_{i,j}U_{\alpha i}
U_{\beta i}U_{\alpha j}U_{\beta j}\sin^2(\phi_{ij}(r))
\end{equation}
is the probability of the neutrino flavor transition
$\nu_\alpha\rightarrow\nu_\beta$ in the case of the CP invariance
and
\begin{equation}
\phi_{ij}(r)=\frac{m^2_i-m^2_j}{2M}r
\end{equation}
is the oscillation phase. The integrand of Eq.~(\ref{intg}) has a
sharp maximum at $r=2t$, and integration can be done approximately
since the width $D$ of the exponential function is much smaller
than that of $D_R(t)$ of the Gaussian one, if $\Gamma t\gg
2d/\lambda_Z$, where $\lambda_Z$ is the Compton wavelength of the
boson ($\Gamma\sim 10^{23}$ s$^{-1}$). The integration is
straightforward, albeit somewhat tedious, and its result is
$$
\frac{dP_{\alpha\beta}}{dr_\nu d\Omega_\nu}\!=\!
W_{\alpha\beta}(2t)(1-\cos\vartheta_\nu+\cos^2\!\vartheta_\nu)
$$
\vspace{-4mm}
\begin{equation}\label{rprob}
\times\frac{3r^2_\nu}{4\pi^{3/2}D^3_R(t)}
\exp\left(-\frac{r^2_\nu}{D^2_R(t)}\right),
\end{equation}
where $\vartheta_\nu$ is the angle between the outgoing neutrino
and the direction of the $Z^0$ boson spin. The probability density
of the oscillation transition
$\bar{\nu}_\alpha\rightarrow\bar{\nu}_\beta$ may be derived in
perfect analogy to above calculations, and has the form
$$
\frac{dP_{\bar{\alpha}\bar{\beta}}}{dr_{\bar{\nu}}
d\Omega_{\bar{\nu}}}\!=\!
W_{\alpha\beta}(2t)(1+\cos\vartheta_{\bar{\nu}}+
\cos^2\!\vartheta_{\bar{\nu}})
$$
\vspace{-4mm}
\begin{equation}\label{rproba}
\times\frac{3r^2_{\bar{\nu}}}{4\pi^{3/2}D^3_R(t)}
\exp\left(-\frac{r^2_{\bar{\nu}}}{D^2_R(t)}\right),
\end{equation}
because $W_{\bar{\alpha}\bar{\beta}}=W_{\alpha\beta}$.

It is seen that the probability densities (\ref{rprob}) and
(\ref{rproba}) involve the amplitude of oscillations depending on
the spatio-temporal localization of a neutrino. It owes its origin
to the evolution of an unstable localized system during its decay.
Obviously, the amplitude can not be found without rigorous
treatment of the decay process as described above. The probability
of transition is obtained by integration of (\ref{rprob}) or
(\ref{rproba}) over a spatial coordinate, and it is equal to
$W_{\alpha\beta}(2t)$.

The detection time $t$ is equal to distance between the detector
and the source, $t=r_\nu=r/2$. Therefore, the oscillation length,
associated with $\Delta m^2_{ij}$, is
\begin{equation}
L_{ij}\!=\!2\pi\frac{M}{|\Delta m^2_{ij}|},
\end{equation}
where the neutrino energy is obviously equal to $M/2$. Two
conditions are necessary for the neutrino oscillations to be
observed. The first condition is the coherence of different mass
eigenstates. Coherence is preserved over distances not exceeding
the coherence length $L_{coh}$. The latter is defined as the
distance at which the phase difference due to energy spreading of
a neutrino, $\delta\epsilon=\Gamma$, obeys the equation
$(\partial\phi/\partial\epsilon)_m\delta\epsilon=2\pi$, which
gives $L_{coh}=L_{osc}Q/\Gamma\sim 10^3L_{osc}$. The second
condition requires that at a distance, which is comparable with
the oscillation length, the spatial separation of the wave packets
corresponding to different mass eigenstates would be considerably
less than the packet width. Such, indeed, is the case
\begin{equation}\label{osl}
\frac{L_{osc}\delta v}{vD}=\frac{\Gamma L_{osc}}{2v^2}
\left|\left(\frac{\partial v}{\partial
m^2}\right)_{\!\!\epsilon}\right|\Delta m^2=\pi\frac{\Gamma}{Q}\ll
1.
\end{equation}
Hence, the spatial separation of the massive neutrino wave packets
does not affect the observation of neutrino oscillations in any
two-particle decay.

All of the above applies to the decay of a single ${\rm Z}^0$
boson. In the case of a steady flux of $Z^0$'s whose decays
provide the incoherent fluxes $I_\alpha$ of the three types of
neutrinos ($\alpha=e,\mu,\tau$), the intensity of neutrino $\beta$
at an any distance from source is proportional to $\sum_\alpha
I_\alpha W_{\alpha\beta}(r)$. Because the probabilities of
emission of different neutrino-antineutrino pairs are equal,
$I_\alpha=I_0$, the sum is equal to $I_0$ according to
Eq.~(\ref{prob}). Hence, the neutrinos (as well as antineutrinos)
do not oscillate, which is in agrement with the result of
Ref.~\cite{Pont}.

{\it Neutrino and antineutrino oscillations in coincidence
measurements.} The term "coincidence" means that the flavors of
both leptons from the same decay are fixed by two detectors. The
flavor-changing process
$\nu_\alpha\bar{\nu}_\alpha\rightarrow\nu_\beta\bar{\nu}_\gamma$
is determined by the probability density
\begin{equation}\label{enp}
\frac{dP_{\alpha\bar{\alpha}\rightarrow\beta\bar{\gamma}}}
{d\mathbf{r}_\nu d\mathbf{r}_{\bar{\nu}}}=
\Bigl|\sum_{i,j}\!U_{\beta i}U^*_{\gamma j}\bigr<\nu_i\bar{\nu}_j|
\Psi(\mathbf{r}_\nu,\mathbf{r}_{\bar{\nu}},t)\Bigr|^2.
\end{equation}
By using Eq.~(\ref{urel}) and (\ref{prob}), one finds
$$
\frac{dP_{\alpha\bar{\alpha}\rightarrow\beta}\bar{\gamma}}
{d\mathbf{r}_\nu d\mathbf{r}_{\bar{\nu}}}=
\Bigl|\Psi_{CM}\Bigl(\frac{\mathbf{r}_\nu\!+
\!\mathbf{r}_{\bar{\nu}}}{2},t\!\Bigr)\Bigr|^2\!
$$
\vspace{-4mm}
\begin{equation}\label{cprob}
\times|F(\theta\varphi)|^2R^2_0(r,t)
W_{\alpha\beta}(r)W_{\alpha\gamma}(r).
\end{equation}
The highest output is obtained when the detectors are located on
the symmetry axis at equal distances from the source
$r_\nu=r_{\bar{\nu}}=t$. For this configuration, the oscillation
amplitude is
\begin{equation}
\frac{(1-\cos\theta)^2\Gamma r^2_\nu}{64\pi^{5/2}D^3_R(r_\nu)}=
\frac{(1-\cos\theta)^2\Gamma}{64\pi^{5/2}r_\nu}
\left(\!\frac{d}{\lambda_Z}\right)^{\!3}.
\end{equation}
This amplitude corresponds to the highest spatial correlation of
the two leptons at a distance equal to $2t$, that is in complete
agreement with the classical picture of a bipartite decay.

The probability of detecting the neutrino $\nu_\beta$ and the
antineutrino $\bar{\nu}_\gamma$ from the decay ${\rm
Z}^0\rightarrow \nu_\alpha+\bar\nu_\alpha$ is obtained by the
integration of Eq.~(\ref{cprob})
\begin{equation}\label{prob1}
W_{\alpha\bar{\alpha}\rightarrow\beta\bar{\gamma}}=\!
\int\!\!dP_{\alpha\bar{\alpha}\rightarrow\beta}\bar{\gamma}=
W_{\alpha\beta}(2t)W_{\alpha\gamma}(2t).
\end{equation}
It is seen that a neutrino and an antineutrino oscillate
independently, and their oscillation lengths are determined by
Eq.~(\ref{osl}). If the pairs $\nu_e\bar{\nu}_e$,
$\nu_\mu\bar{\nu}_\mu$ and $\nu_\tau\bar{\nu}_\tau$ are emitted
independently, and their fluxes are equal, then the probability of
observing $\nu_\beta$ in one detector and $\bar{\nu}_\gamma$ in
another one will be equal
\begin{equation}
W_{\beta\bar{\gamma}}= \sum_\alpha
W_{\alpha\beta}W_{\alpha\gamma},
\end{equation}
which is different from Eq.~(\ref{prob1}) for the decay of a
single boson. For the detectors connected to a coincidence circuit
with one output, the oscillation pattern is distorted due to
mixing of signals from two detectors. In order to observe the
independent oscillations of neutrinos and antineutrinos, it is
necessary to synchronize the time on two detectors to make sure
that the leptons come from one decay.

In summary, an accurate analytical solution for the joint quantum
state of flavor neutrino and antineutrino following the decay of a
${\rm Z}^0$ boson has been found. The evolution of the state
provides yet another exactly calculable illustration of the famous
Einstein-Podolsky-Rosen thought experiment. The new effect is the
entanglement between two oscillating leptons. It is shown that the
oscillations of neutrino and antineutrino from the same decay
proceed independently and may be observed by measuring the flavor
of both particles. Massive neutrinos and antineutrinos are not
emitted in a state with definite momentum, however the energy of
the two-lepton wave packet is fixed and is equal to the decay
energy. Hence it can be argued that the neutrino mass eigenstates
produced in bipartite decay have the same energy.

Despite its purely academic interest, the solved problem is
essential to the understanding of the physics of neutrino
oscillations. Besides, the sources of neutrinos from the ${\rm
Z}^0$-decay, as well as detectors recording neutrinos from a
single decay may well be conceivable in the future experiments.
Accordingly, I have considered the experiments allowing to observe
oscillations of one lepton as well as both leptons from the decay
of a single boson. If three neutrino-antineutrino pairs are
emitted independently, the oscillations can be observed also by
the two detectors connected into the coincidence circuit with one
output or synchronized in time. A similar result was obtained in
Ref.~\cite{Zat}. However, the results of that study are based on
the erroneous assumption that $Z^0$ boson decays into a coherent
superposition of neutrino-antineutrino pairs with different
flavors \cite{Ivn}.

The work was supported by the Grant NS-215.2012.2 from the Russian
Ministry of Education and Science.


\begin{thebibliography}{99}
\bibitem{Pav1} The rigorous substantiation of the far zone
approximation has been given in the work of the author \cite{Pav}.
\end{thebibliography}
\end{document}